\documentclass[twocolumn,epjc3]{svjour3} 
\usepackage[utf8]{inputenc}
\usepackage{amssymb}
\usepackage{amsmath}
\usepackage{comment}
\usepackage{lineno}
\usepackage{comment}
\usepackage{overpic}
\usepackage{wasysym}
\RequirePackage{graphicx,subfig,multirow,ulem}

\DeclareGraphicsExtensions{.pdf}
\RequirePackage[numbers,sort&compress]{natbib}
\RequirePackage[colorlinks,citecolor=blue,urlcolor=blue,linkcolor=blue]{hyperref}

\newcommand{\CR}{$^{51}$Cr}
\newcommand{\CENNS}{CE$\nu$NS}

\title{Coherent elastic nuclear scattering of \CR\ neutrinos}
\author{
C. Bellenghi\thanksref{UNISAP,INFN-RM1}, 
D. Chiesa\thanksref{MIB,INFN-MIB}, 
L. Di Noto\thanksref{UNIGE,INFN-GE}, 
M. Pallavicini\thanksref{UNIGE,INFN-GE}, 
E. Previtali\thanksref{MIB,INFN-MIB} and
M. Vignati\thanksref{INFN-RM1, e1}
}
\thankstext{e1}{e-mail: marco.vignati@roma1.infn.it}

\institute{
{Dipartimento di Fisica, Sapienza Universit\`{a} di Roma, Roma I-00185 - Italy}\label{UNISAP}
\and
{INFN Sezione di Roma, Roma - Italy}\label{INFN-RM1}
\and
{Dipartimento di Fisica, Universit\`{a} di Milano - Bicocca, Milano I-20126 - Italy}\label{MIB}
\and
{INFN Sezione di Milano Bicocca, Milano - Italy}\label{INFN-MIB}
\and
{Dipartimento di Fisica, Universit\`{a} di Genova, Genova I-16146 - Italy}\label{UNIGE}
\and
{INFN  Sezione di Genova, Genova - Italy}\label{INFN-GE}
}

\begin{document}

\maketitle

\begin{abstract}
Searches for new physics in the coherent elastic neutrino-nucleus scattering require a precise knowledge of the neutrino flux and energy spectrum.
In this paper we investigate the feasibility and the performance of an experiment based on a \CR\ source, whose neutrino spectrum is known and whose activity can be heat-monitored at few permil level.
With a 5~MCi source placed at $\sim25$~cm  from the detector, under an exposure of two \CR\ half-lives (55.4 days), we evaluate 3900 (900) counts on a 2000~cm$^3$  target of germanium (sapphire) featuring an energy threshold of 8 (20)~eV. To further increase the exposure, multiple activations of the same source could be possible.
\end{abstract}

\section{Introduction} 
Predicted in 1973~\cite{Freedman:1973yd} and observed for the first time in 2017~\cite{Akimov:2017ade}, the coherent elastic neutrino-nucleus scattering (CE$\nu$NS)
has been  proposed to search for neutrino magnetic moment, Z' exchange,  non-standard interactions,  sterile neutrinos~\cite{Dodd:1991ni, Barranco:2005yy, Formaggio:2011jt, Dutta:2015nlo, Lindner:2016wff},
and to perform precision measurements of the nuclear form factor~\cite{Patton:2012jr} and of $\sin^2\theta_W$ at low momentum transfer~\cite{Erler:2017knj}.
The advantage of CE$\nu$NS  relies in the relatively large neutrino
cross section, which scales quadratically with the number of neutrons in the nucleus~\cite{Freedman:1977xn}:
\begin{equation}\label{eq:xsec}
\frac{d\sigma}{dT} = \frac { G_F^2} {4\pi}F^2(q^2)  Q_W^2 M_A\left(1-\frac{M_A T}{2 E^2_\nu}\right) 
\end{equation}
where $G_F$ is the Fermi constant, $F$ the nuclear form factor, $Q_W = N-Z (1-4\sin^2 \theta_W)\simeq N$, $N$ the number of neutrons in the nucleus, $Z$ the number of protons, $\theta_W$ the Weinberg angle, $M_A$ the nucleus mass, $T$ the nucleus recoil energy and $E_\nu$ the neutrino energy. 
On the other hand, the recoil energy
$T \leq 2 E_\nu^2/M_A$
is small, of the order of  tens or hundreds of eV for MeV  neutrinos on a Germanium target.
To achieve small energy thresholds, germanium diodes~\cite{Barbeau:2007qi, Hakenmuller:2019ecb, Kerman:2016jqp}, CCD's ~\cite{Aguilar-Arevalo:2016khx,Tiffenberg:2017aac}, and cryogenic
phonon detectors of silicon, germanium~\cite{Agnolet:2016zir,Billard:2016giu}, sapphire and CaWO$_4$~\cite{Strauss:2017cuu} have been proposed. The most competitive thresholds have been
demonstrated by  phonon detectors (tens of eV) and  CCDs (1 electron).

Current searches for new physics in \CENNS\ could be limited by systematic errors on the neutrino flux and energy spectrum, as in the case of neutrino spallation sources, where the flux normalization precision is currently $\sim10\%$~\cite{Akimov:2017ade}, and power nuclear reactors, where models of the neutrinos emitted by the nuclear fuel limit the precision to several percent ~\cite{Hayes_reactor_flux,Vogel_reactor_flux}. 
Instead, artificial nuclear sources can be characterized with high precision and were already employed in the past in the GALLEX (\CR)~\cite{GallexPLB} and SAGE (\CR, $^{37}$Ar) ~\cite{SAGE} experiments, and recently proposed by the SOX experiment ($^{144}$Ce)~\cite{SOX} which, however, was stopped by difficulties occurred during the isotope mass production.

\begin{figure}[tb]
    \centering
    \includegraphics[width=0.45\textwidth]{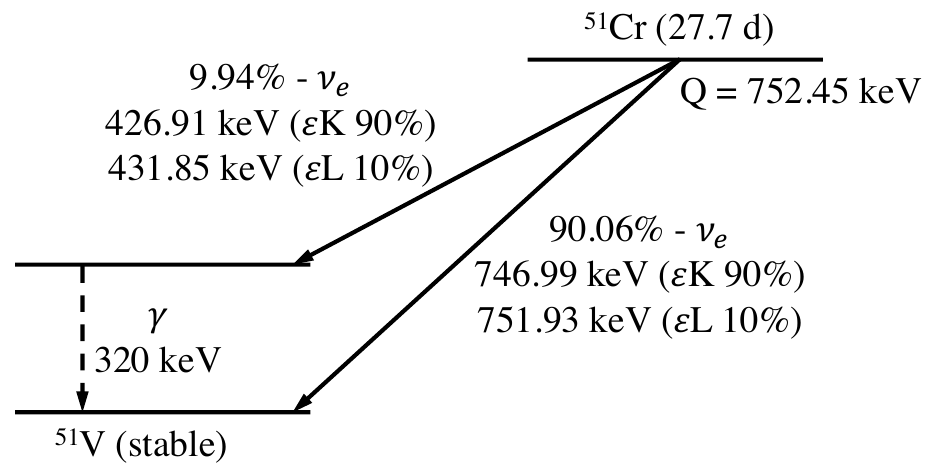}
    \caption{\CR\ decay scheme~\cite{DecayScheme}. }
    \label{fig:decayscheme}
\end{figure}

In this paper we set the requirements for  an experiment on CE$\nu$NS with a high-intensity  source of \CR,
an electron-capture decaying isotope with a half-life of 27.7 days. The neutrino spectrum consists of four monochromatic lines, including
the most energetic 747 keV (81\%) and 752 keV (9\%)  lines which can be exploited for the coherent elastic neutrino-nucleus observation (see the complete decay scheme in Fig.~\ref{fig:decayscheme}).
One advantage of this choice consists in the possibility of reusing the source of the GALLEX experiment, which is still owned by INFN: 36~kg of Cromium 38.6\% enriched in $^{50}$Cr, that need to be activated at a nuclear reactor. Another advantage consists in the possibility of measuring the activity at few per mill level with small modifications of the thermal calorimeter already developed by SOX for its $^{144}$Ce source~\cite{calorimetro}. 
Additionally,  the detection of neutrinos produced by the electron capture could provide complementary information to the detection of anti-neutrinos  from nuclear reactors. 

After reviewing the possibilities to activate the source, studying its gamma-ray shield and the activity measurement, we propose an experimental configuration placing the source as close as $\sim 25$~cm to the detector. Given that the maximum recoil energy amounts to tens of eV (Fig.~\ref{fig:xsec}), only the case of cryogenic phonon detectors is considered. CCDs are also highly promising, provided that the energy conversion efficiency
of nuclear recoils to electrons is precisely measured.

\section{Activation of the Cr sample} 
We take as reference the irradiation performed in 1994 to produce the 1.7~MCi source of  GALLEX~\cite{GallexPLB,GallexSourceNIM}. 
This source was obtained by neutron activation of  the full sample of 36~kg of Cr, in the form of  $\sim$1~mm$^3$ metal chips, with a total number of $^{50}$Cr nuclei amounting to $1.6\times10^{26}$. The irradiation was performed at the Silo\'e reactor in Grenoble, nowadays decommissioned, and lasted  23.8 days. 
The reactor core was specially reconfigured to host zircalloy containers for the Cr chips, that were exposed to an average thermal neutron flux of $5.2\times10^{13}$~n/cm$^{2}$s. This number was directly measured during the activation and takes into account the perturbation of the neutron flux due to the self-shielding of the Cr sample. The value of the  unperturbed neutron flux is not available in literature however we can refer to the IAEA's Research Reactor Database~\cite{RRDB}, where the maximum total flux of the Silo\'e reactor is quoted in the range $2-4\times10^{14}$~n/cm$^{2}$s.

In the following we survey some of the possible sites to activate the source.
Power reactors are not suitable, because they have no internal sites to place irradiation targets. Conversely, some research reactors have facilities which in principle are suitable in terms of neutron flux  and space available to fit the Cr target. We restrict the list to sites with high neutron fluxes, in order to activate the largest possible number of nuclei.
\begin{itemize}
\item[-]{The BR2 reactor at Mol (Belgium), is not only one of the most powerful (up to 100~MW) material testing reactor in the world, but also a major facility for radioisotope production. The reactor, whose core is 80~cm high, has a  20~cm~$\diameter$ {\it central large channel},  with maximum thermal neutron flux $10^{15}$~n/cm$^{2}$s, and 3 {\it peripheral large channels} with $3\times10^{14}$~n/cm$^{2}$s. The  typical cycle is three weeks~\cite{RRFM2017}}.
\begin{figure}[t]
    \centering
   \includegraphics[width=0.45\textwidth]{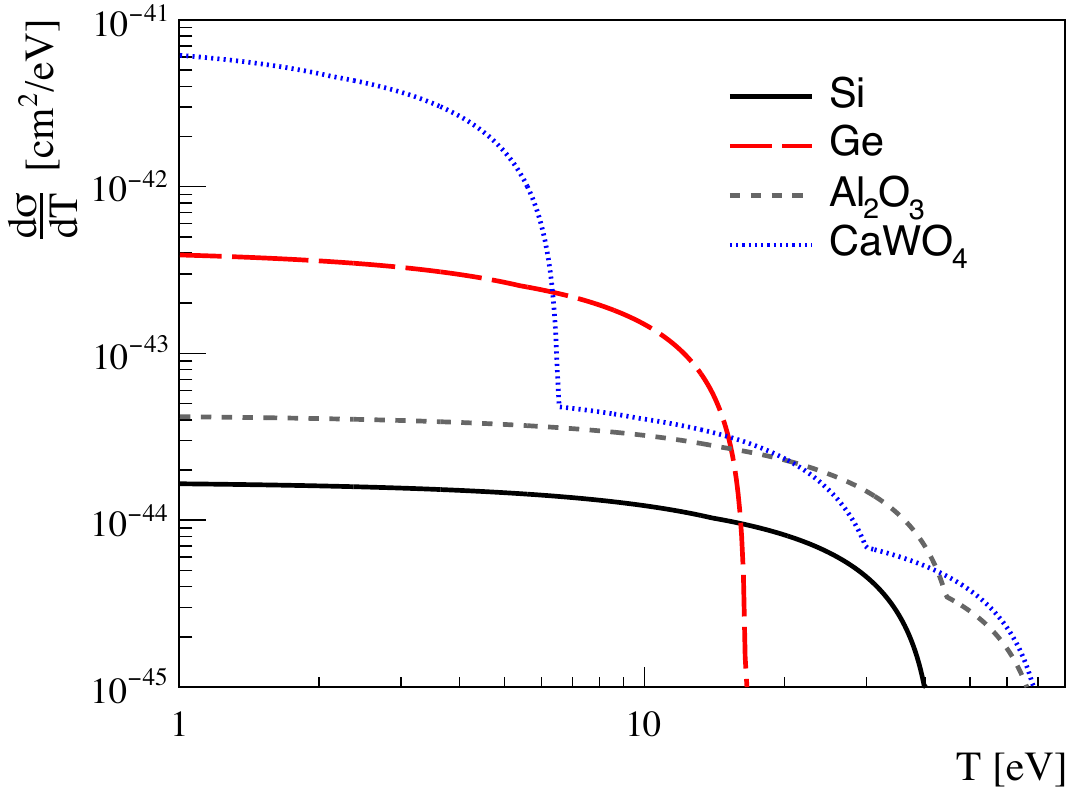}
    \caption{Differential cross section as a function of recoil energy $T$ for \CR\ neutrinos on typical targets used in cryogenic experiments: silicon, germanium, sapphire and calcium tungstate. The nuclear form factor in Eq.~\ref{eq:xsec} is assumed unitary~\cite{AristizabalSierra:2019zmy}.}
    \label{fig:xsec}
\end{figure}

\item[-]{The High Flux Isotope Reactor (HFIR) at Oak Ridge (USA), 85~MW power, has irradiation positions with thermal fluxes up to 1.2--2.5$\times10^{15}$~n/cm$^{2}$s,
however irradiations must be carefully reviewed with respect to their neutron poison content, which is limited because of their adverse effects on the reactor power distribution and fuel cycle length~\cite{HIFR}. Nevertheless HFIR  has also 16 small (4~cm~$\diameter$) and 6 large (7.2~cm~$\diameter$) \textit{vertical experiment facilities} (VXF), 50~cm high, with maximum thermal neutron flux in the range 4--7$\times10^{14}$~n/cm$^{2}$s. Large neutron poison loads in VXF facilities are of no particular concern~\cite{konings2011}.
}

\item[-]{Other reactors with similar features are the Advanced Test Reactor (ATR) at Idaho National Laboratory (USA), the HFR reactor at Petten (Netherlands), and the SM3 or MIR.M1 reactors at Dimitrovgrad (Russia). The Jules Horowitz Reactor (JHR), currently under construction in Cadarache (France) will feature facilities in the reflector region with thermal fluxes up to 5$\times10^{14}$~n/cm$^{2}$s~\cite{JHR}.
}
\end{itemize}
Considering an average unperturbed thermal neutron flux of $5\times10^{14}$~n/cm$^{2}$s, we calculated the activity that would be obtained after 24~days of irradiation of the Cr target.
We performed Monte Carlo simulations with the MCNP code~\cite{mcnp}  to evaluate the self-shielding effect in different irradiation geometries and obtained an activity in the range of 3.5--7~MCi. 

In the following we assume the activity of  $^{51}$Cr as 5~MCi.
If not sufficient, the source could be re-activated several times,  since the number of $^{50}$Cr isotopes transmuted would correspond to the $0.4$\% only of the GALLEX sample.
Increasing the irradiation time instead is not obvious, because irradiation cycles of research reactors typically last three weeks.

\section{Source shielding and activity measurement} 

$^{51}$Cr  produces only 320~keV gammas in 10\% of the decays that, from the safety point of view,   can be easily  shielded by a few centimeters of tungsten.
The most worrisome  radiation, instead, comes from the  sample impurities.
In GALLEX, despite the special care taken in all the chemical processing stages, impurities with (n,$\gamma$) cross section in the barn range were activated during irradiation, producing radionuclides with lifetime of several days. In particular, the 1.3-1.5 MeV $\gamma$ emitter $^{110m}$Ag  ($T_{1/2}=150$~d) was found with an activity of 4 GBq~\cite{Gallex_impurities}, while  other long-lived emitters such as  $^{46}$Sc (1.2~MeV), $^{60}$Co (1.1-1.3~MeV) and $^{124}$Sb (2~MeV) were found with an activity smaller than 0.5~GBq~\cite{GallexSourceNIM}. 
Assuming the same activation factor of GALLEX, an activity around 12 GBq is expected from a 5 MCi source.  To comply the safety rules (dose at contact $<$ 100-200 uSv/h),
an attenuation factor $>2000$ is necessary, which could be achieved with a shield of tungsten at least 8 cm thick.

The $\gamma$s escaping the shield may reach the detector and generate a continuum background down to the signal region via Compton scattering. Aside the impurities,  the $^{51}$Cr would produce  $\gamma$s up to $\sim 750$ keV  via inner bremsstrahlung which, even if suppressed by a factor $7\times10^{-4}$,  is still abundant in a 5 MCi source (185~PBq). 
The purification level of the source and the shield thickness needed to not overwhelm the neutrino signal depend on the geometry of the experiment and on the detector materials, and are analysed in the next section.

Since the 320 keV $\gamma$, the X and Auger radiations are completely absorbed and the heat released by impurities is negligible, a total of (36.51$\pm$ 0.161) keV is liberated per each $^{51}$Cr decay, corresponding to 422~W for a 5 MCi source.
To measure the activity, just before and after the data taking, the shielded source can  be inserted inside a thermal heat exchanger in which ultra-pure water circulates and eventually absorbs the heat, increasing its temperature. 
As it has been shown in Ref.~\cite{calorimetro} for the case of $^{144}$Ce, the thermal losses due to the convection, conduction and irradiation can be minimized at negligible levels. By measuring  the mass flow and the inlet and outlet water temperatures, the heat power can be evaluated with 0.2\% precision. 
Since the $^{51}$Cr half-life is about 10 times shorter than the $^{144}$Ce half-life for which the apparatus  was designed and calibrated, a systematic shift of about 0.5\% in the measured power arises, due to the delay time induced by the heat propagation from the source to the water. Nevertheless the delay time can be decreased by improving the thermal contact between the shielded source and the heat exchanger, or it can be measured during the calibration phase, allowing the precision to be preserved at few per mill level.

\section{Prototype experiment} 

Next generation phonon detectors for dark matter and \CENNS\ search aim at kg detectors with energy threshold of tens of eV. 
An energy threshold of 20~eV has been already demonstrated on a 0.5~g ($5\times5\times5$~mm$^3$) sapphire target~\cite{Strauss:2017cam}, while
60~eV have been demonstrated on a 33.4~g (20~mm $\diameter\times20$~mm) germanium target~\cite{Armengaud:2019kfj}. 
The challenge consists in scaling up the technology to thousands detectors, by improving the
detector reproducibility and the multiplexing capability. While it is customary to quantify the size of the detector by its mass, here we choose to fix its volume, since in cryogenic experiments the technology scale up depends mainly on the size and number of detectors, irrespective of their density, and since in the setup we are proposing the neutrino flux depends on the detector geometry. We choose a volume of 2000 cm$^3$, which corresponds to $\sim 5-12$~kg depending on the target considered, and we choose a cylindrical shape with 10~cm diameter.

\begin{figure}[t]
    \centering
    \includegraphics[width=0.48\textwidth]{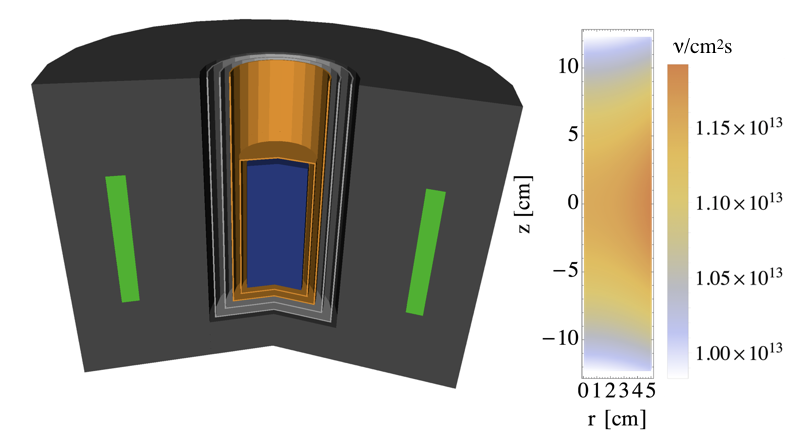}
    \caption{Left: cylindrical detector of 2000~cm$^3$ with 5 cm radius (blue), placed inside a cylindrical cryostat 5~cm thick (orange - light gray). The 10000~cm$^3$ source (green) is deployed in a toroid around the cryostat and shielded in the radial direction by 12~cm of tungsten (dark gray). The source is at an average distance of 25 cm from the detector center, and its height (25 cm) is chosen to maximize the neutrino flux. Right: simulation  of the neutrino flux at the detector in cylindrical coordinates for a \CR\ source with initial activity of 5~MCi averaged over 2 half-lives (55.4 days).}
    \label{fig:simulation}
\end{figure}

To maximize the neutrino flux we place the source as close as possible to the detector. Custom cryostat tails can be realized with a thickness of $\sim 5$~cm between the low temperature volume hosting the detector and the outside. The source, surrounded by its shield, is then placed around the cryostat, in a toroidal configuration. We perform simulations to determine the source height and thickness that maximize the neutrino flux, keeping the volume fixed to the GALLEX measured value of 10000~cm$^3$. To determine the shield thickness we also simulate the irreducible $\gamma$ background from the inner bremsstrahlung of $^{51}$Cr, restricting to Sapphire targets. 
Figure~\ref{fig:simulation} sketches the proposed configuration of the experiment, where we also allow for 1 cm tolerance between the detector and the inner cryostat shield and between the outer cryostat shield and the source shield. The internal thickness of the tungsten shield that ensures a background one order of magnitude below the neutrino signal is 12~cm. The external thickness is fixed to the safety requirement of 8~cm.
The source height and thickness that maximize the neutrino flux are 25 and 2.5~cm, respectively.

\begin{figure}[t!]
    \centering
    \includegraphics[width=0.5\textwidth]{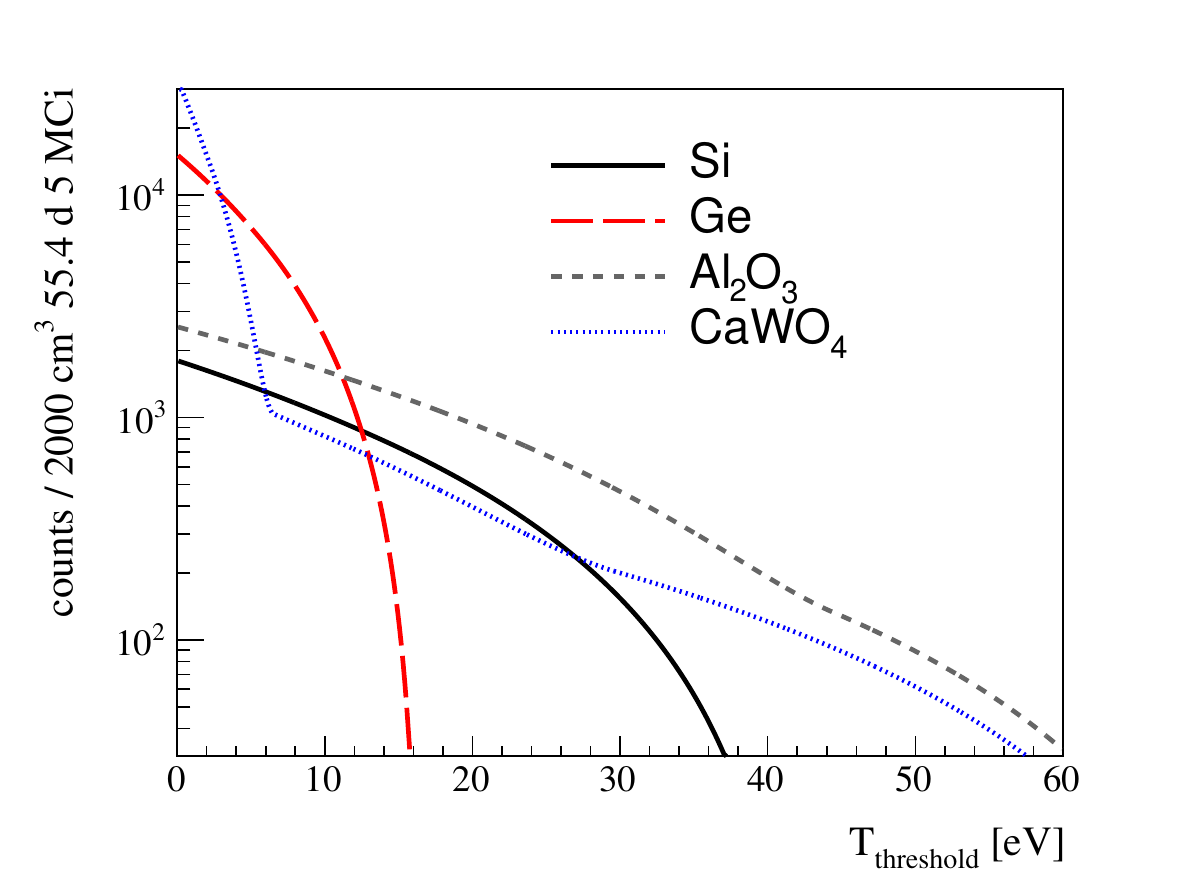}
    \caption{Counts above threshold for different types of 2000 cm$^3$ detectors from 55.4 days of exposure to a source of 5~MCi initial activity, in the same experimental configuration of Fig.~\protect\ref{fig:simulation}.}
    \label{fig:counts}
\end{figure}

The neutrino flux at the detector, averaged over two \CR\ half-lives, ranges between $1.0$ and $1.2\times10^{13}\nu /{\rm cm^2 s}$ depending on the impact position, with an average of $1.1\times10^{13}\nu /{\rm cm^2 s}$.
The counts registered in the detector as a function of its energy threshold, for the targets used in present experiments, are shown in Fig.~\ref{fig:counts}. With a threshold of 20~eV, sapphire detectors would register 900 counts, corresponding to a 3.3\% precision on the cross-section measurement without considering backgrounds. In case the detector technology advances and pushes down the threshold, a germanium target would register 3900 counts with a threshold of 8 eV, corresponding to a 1.6\% precision. Silicon would provide a factor 2 lower counts than sapphire whereas calcium tungstate, to overcome the germanium performance, would require an energy threshold below 3 eV, a value that appears difficult to achieve. 

 The background at low energies due to Compton scattered $\gamma$s must be studied with detailed measurements and simulations. With our Geant4-based simulation  we estimated that, with a 12~cm tungsten shield, Ag impurities in the GALLEX sample must be reduced from ppm to ppb to not generate a background\footnote{
The Geant4 simulation we use is not trustworthy at energies below 1 keV. We extrapolate the continuum generated by Compton scattering in the region $1-10$~keV down to the neutrino signal region, which is below 60 eV. Dedicated measurements are needed to measure and simulate the physical processes taking place at such low energies.}.
The environmental background is hard to predict since there are no measurements at such low energies: experiments will hopefully start soon, and provide key information. The most worrisome background could be due to neutrons, which could be absorbed with additional shielding. 

To increase the exposure, aside the increase of the detector size,  rerunning the same experiment is in principle possible. As described in Sec. 2,  the Cr sample could be exposed back to neutrons and the experiment could restart. In this view, and also considering the short half-life, the experiment should be performed close to the nuclear reactor complex where the source is activated.  For a 100~MW reactor, neutrinos from the core would  appear as a continuum background 10 times smaller than the signal if the experiment is located at a distance greater than 25~m.

\section*{Acknowledgements}
We thank Andrea Borio di Tigliole and Angelo Cruciani for useful discussions.
This work was partially supported by the European Research Council (FP7/2007-2013) under Contracts CALDER No. 335359 and SOX No. 320873 and makes use of the Arby software for Geant4 based Monte Carlo simulations, which has been developed in the framework of the Milano -- Bicocca R\&D activities and is maintained by O. Cremonesi and S. Pozzi.

\bibliographystyle{spphys.bst} 
\bibliography{51CrCENNS}


\end{document}